\documentclass[11pt,preprint]{aastex}
\usepackage{apjfonts}
\usepackage{epsfig}
\usepackage{amsmath}

\begin{document}
\title{Modeling the gamma-ray emission in the Galactic Center with a fading cosmic-ray accelerator}

\author{Ruo-Yu Liu\altaffilmark{1}, Xiang-Yu
Wang\altaffilmark{2,3,1}, Anton Prosekin\altaffilmark{1},
Xiao-Chuan Chang\altaffilmark{2,3}}
\altaffiltext{1}{Max-Planck-Institut f\"ur Kernphysik, 69117
Heidelberg, Germany; ruoyu@mpi-hd.mpg.de} \altaffiltext{2}{School
of Astronomy and Space Science, Nanjing University, Nanjing
210093, China; xywang@nju.edu.cn} \altaffiltext{3}{Key laboratory
of Modern Astronomy and Astrophysics (Nanjing University),
Ministry of Education, Nanjing 210093, China}

\begin{abstract}
{Recent HESS observations of the $\sim200$~pc scale diffuse
gamma-ray emission from the central molecular zone (CMZ) suggest
the presence of a PeV cosmic-ray accelerator (PeVatron)  located
in the inner $10$~pc region of the Galactic Center. Interestingly,
the gamma-ray spectrum of the point-like source (HESS~J1745-290)
in the Galactic Center shows a cutoff at $\sim 10$~TeV, implying a
cutoff around $100$~TeV in the cosmic-ray proton spectrum. Here we
propose that the gamma-ray emission from the inner and the outer
regions may be explained self-consistently by run-away protons
from a single, yet fading accelerator. In this model, gamma rays from
the CMZ region are produced by protons injected in the past, while
gamma rays from the inner region are produced by protons injected
more recently. We suggest that the blast wave formed in a tidal
disruption event (TDE) caused by the supermassive black hole (Sgr A*)
could serve as such a fading accelerator. With typical parameters
of the TDE blast wave, gamma-ray spectra of both the CMZ region
and HESS~J1745-290 can be reproduced simultaneously.
Meanwhile, we find that the cosmic-ray energy density profile in
the CMZ region may also be reproduced in the fading accelerator
model when appropriate combinations of the particle injection
history and the diffusion coefficient of cosmic rays are adopted.
}
\end{abstract}

\section{Introduction}
Recently, the HESS collaboration reported the deep gamma-ray
observations with arcminute angular resolution of the central
molecular zone (CMZ) surrounding the center of our Galaxy
\citep{HESS_GC16}, extending out to $r\sim 250$~pc and $\sim
150$~pc at positive and negative Galactic longitudes respectively.
The brightness distribution of very high energy (VHE) gamma rays
shows a strong correlation with the locations of massive gas-rich
complexes. This discovery, combined with the fact that in the leptonic
scenario severe radiative losses would be expected for multi-TeV
electrons in the Galactic Center, points to a hadronic origin of
the diffuse VHE emission. The inferred radial profile of $>10$~TeV
cosmic-ray (CR) energy density is consistent with $1/r$ dependence
in the entire CMZ region,  and it was argued for a centrally located
source  in the inner $\sim 10$~pc region with a quasi-constant
injection operating over at least
$1000$~years, given the diffusive propagation of cosmic rays
\citep{HESS_GC16}. The best-fit gamma-ray spectrum shows a single
power-law form with a photon index of $\sim 2.3$ extending to
energies larger than tens of TeV without a break or a cutoff,
which indicates the existence of a PeV proton accelerator, or the
so-called "PeVatron".

Interestingly, the location of the PeVatron appears to coincide
with the  central point-like gamma-ray source HESS~J1745-290,
which shows a harder spectrum with a photon index of $\sim 2.1$
and, however, a clear cutoff at $\sim 10$~TeV \citep{HESS_GC09}.
The GC has long been suggested as a cosmic-ray proton accelerator
\citep[e.g.][]{Ptusin81, Said81, Aharonian05b, Liu06, Fujita16}
and a hadronic origin was also suggested for the gamma-ray
emission of HESS~J1745-290 \citep{Aharonian05b, Chernyakova11},
while the leptonic origin may work well too \citep{Aharonian05a,
Hinton07}. There is no obvious connection between HESS~J1745-290
and the gamma rays from the outer CMZ region. However, as protons
accelerated at the center source will produce gamma rays via the
proton--proton collision with the gas along the path to the outer
CMZ region, they may potentially also explain the gamma-ray
emission from HESS~J1745-290 \citep{Aharonian05b, HESS_GC16}. In
this scenario, some additional mechanisms must be
introduced to reconcile the 10\,TeV cut--off in the spectrum of
HESS~J1745-290 with the extension of the proton spectrum to PeV
energies in the outer CMZ region. One possible solution is
invoking the absorption due to a dense infrared photon field near
the GC \citep{HESS_GC16}. However, based on the current infrared
data of the GC,  the attenuation of the multi TeV photon flux
originating from the GC is probably not strong enough to cause a
cutoff \citep{Moskalenko06, Zhang06, HESS_GC09, Celli16},
though an extremely clumpy distribution
of the infrared photons might overcome this difficulty
\citep{Guo16}.

In this work, we propose a fading proton accelerator located at the inner
$10$~pc region as the origin of the gamma-ray emission from both
HESS~J1745-290 and the CMZ. In this model, the source
was more powerful in the past and injected protons beyond PeV
energies. These protons have propagated to the outer CMZ
region, where, at the present time, they produce the gamma-ray
emission with the spectrum extending to high energies. On the other hand,
as the source power gradually fades and the maximum  energy of the
injected protons declines, the gamma-ray emission from the inner region,
which arises from recently injected protons, develops a cutoff in
the spectrum, which decreases towards the Galactic Center. We will show
that this model explains simultaneously the gamma-ray spectra of the CMZ
region and the inner point-like source.

As the particle injection rate of a fading accelerator decreases with time as well,
it is apparently inconsistent with the suggestion by \citet{HESS_GC16} that the CR
energy density profile indicates a constant injection rate assuming a
spatially independent diffusion coefficient. So we need to check
whether the measured CR density profile can be reproduced in the
fading accelerator model.

The rest part of this paper is organized as follows. We fit the gamma-ray spectral
data of both the inner and outer region simultaneously in a specific scenario of
fading accelerator, i.e., a tidal disruption event
by the supermassive black hole in the Galactic Center drives a
blast wave which accelerates cosmic rays (\S 2). In \S 3, we study
the energy density profile of cosmic rays in the general fading
accelerator model, paying particular attention to the influence of
the particle injection history and the diffusion coefficient.
The discussion and conclusion are given in \S 4.


\section{Modeling the gamma-ray spectra of the inner and outer regions}

\subsection{TDE blast wave as a fading particle accelerator}
As the cosmic-ray acceleration process and injection history
depends on the details of the accelerator, we now look for
specific astrophysical sources that can act as such a fading
accelerator and study whether the  spectra of the inner and
outer regions can be produced simultaneously in this scenario.
\cite{HESS_GC16} already pointed out that supernova remnants may
not be responsible for the CRs, because the maximum proton energy
will drop below PeV too quickly for a single SNR, while a
high supernova rate in the central 10\,pc seems unlikely. Limited
by the position of the accelerator, \cite{HESS_GC16} suggested the
supermassive black hole at the GC (Sgr~A*) as the most plausible
source for these CRs among several potential candidates.

We propose that a tidal disruption event (TDE) by the supermassive
black hole at the GC could be such a fading accelerator. When a
star gets too close to the supermassive black hole, it will be
disrupted by the tidal force of the black
hole \citep[e.g.][]{Hills75, Carter82, Rees88}, leading to a
transient accretion disk and sometimes also a relativistic
outflow. There are growing number of candidate TDEs occurred in
other galaxies that have been discovered in X-ray, ultraviolet,
and optical surveys (see \citealt{Komossa15} for a review).
Three TDE candidates have been also detected  in non-thermal  X-ray  and radio
emission, i.e., Swift J1644+57 \citep{Bloom11}, J2058+05 \citep{Cenko12_TDE}, and J1112-8238 \citep{Brown15}.
The non-thermal X-ray  and radio emissions are thought
to be produced by relativistic jets, in which shocks occur and
accelerate non--thermal electrons. The isotropic radiation
energies in X-rays in all three jetted TDEs are about
$3\times10^{53}$~erg, so the isotropic kinetic energy of the jets
are of the order of $10^{54}$~erg. Assuming an opening angle  of
$\theta_j\sim0.1$, the real jet energy is about $10^{52}$~erg. The
event rate of TDEs caused by supermassive black holes of mass
$10^{6}-10^7{M_\odot}$ is about $10^{-5}-10^{-4} {\rm yr^{-1}}$
per galaxy \citep{WangJX04, Stone16}. Interestingly, it has
been suggested that TDEs in the GC can power the diffuse gamma-ray
emission in the GC via hadronic interactions \citep{Cheng07, ChenX16}.
We here suggest that the TDE blast wave is a fading accelerator
and it can explain both gamma-ray emission spectrum and the CR
density profile in the GC. In the calculation below, we use eV as
the unit of particle energy and c.g.s units for other
quantities. We denote by $Q_x$ the value of the quantity $Q$ in
units of $10^x$ unless specified.

The TDE jet expands sideways when its Lorentz factor drops
to $\Gamma\la 1/\theta_j$, and the blast wave approaches
spherical symmetry soon after it enters the sub-relativistic phase
\citep{Livio00}. Later on, being decelerated further by the
ambient medium, the blast wave enters the non-relativistic phase
(the Sedov phase) at
\begin{equation}
t_{s}=[3E/(4\pi n m_p c^5)]^{1/3}=0.06 \,E_{52}^{1/3}n_4^{-1/3}\,{\rm yr}\,
\end{equation}
where $n$ is the number density of gas in the GC region. This time
is shorter than one year for typical parameters. In the Sedov
phase, the speed of the blast wave (in units of the light speed) is
\begin{equation}
\beta=\left(\frac{12E}{125\pi n m_p c^5 t^3
}\right)^{1/5}=4.7\times10^{-4}\; E_{52}^{1/5}n_4^{-1/5}
\left(\frac{t}{\rm 10^4 yr}\right)^{-3/5}.
\end{equation}
The radius of the blast wave is then
\begin{equation}
R=\frac{5}{2}\beta c t=3.6\; E_{52}^{1/5}
n_4^{-1/5}\left(\frac{t}{\rm 10^4 yr}\right)^{2/5}\, {\rm pc}.
\end{equation}
To estimate the maximum CR proton energy accelerated by
the blast wave, we adopt the formula given by \citet[][also see \citealt{Bell15}]{Bell13}, which considers
the amplification of magnetic field by the non-resonant hybrid instability, i.e.,
\begin{equation}\label{eq:Emax}
E_{p,\rm max}=55\,\eta_{0.1}E_{52}^{3/5} n_4^{-1/10}\left(\frac{t}{\rm 10^4\, yr}\right)^{-4/5}\,{\rm TeV}.
\end{equation}
Note that here we have already substituted the evolution of shock
velocity and  shock radius into the original equation.
$\eta_{0.1}$ is the acceleration efficiency normalized to
0.1\footnote{The value of $\eta$ is estimated to be $0.03$ in
\cite{Bell13}, assuming that only CRs of the highest-energy
decade escape and drive the instability. However, gamma-ray
observations of many CR-illuminated molecular clouds indicate a
broad--spectrum escape of CRs from supernova remnant
\cite[e.g.,][see also \citealt{Malkov13}]{HESS08_w28, Fermi10_w28,
Uchiyama12, Fermi13_SNR}, rather than a high--energy--biased
escape. Since the growth rate of the instability is proportional
to the escaping CR current, consideration of a broad-spectrum escape of
CRs could increase this the efficiency by an order of magnitude.}.
{The time--dependent behavior of the maximum energy $E_{\rm max}$
is important for the spectral fitting. If $E_{\rm max}$ decreases
faster, the two cutoff energies in the respective spectra of the
inner region and the CMZ region would have larger difference. So
that the spectrum of the CMZ region can extend up to higher
energies while the spectrum of the inner region keeps the same
cutoff energy.}

The total luminosity of cosmic rays in the Sedov phase can be
estimated as \citep{Gabici11}
\begin{equation}
L_p(t >1\,{\rm GeV})=f_{CR} \frac{nm_pu_{\rm sh}^3}{2}4\pi
R^2=3.6\times 10^{40} \,f_{\rm CR} E_{52}\left(\frac{t}{10^4 \rm yr}\right)^{-1} \,{\rm erg/s} ,
\end{equation}
where $u_{\rm sh}=\beta c$ is the velocity of the blast wave and
$f_{\rm CR}$ is the  fraction of the kinetic energy of the blast
wave that goes into injected CRs. $f_{\rm CR}$ is at the
level of $\sim 1\%-10\%$ for the blast waves of supernova
remnants, according to the gamma-ray analysis on our Galaxy
\citep{Strong10} and nearby starburst galaxies \citep{Paglione12,
Peng16}. For simplicity, we use a time-independent $f_{\rm
CR}$.

{We note that the dynamic evolution of the blast wave driven by
TDE should be  identical to that of the supernova remnant except
for the one-order-of-magnitude larger kinetic energy, since
both follow the Sedov solution. This larger kinetic energy
leads to a $E_{p, \rm max}$ larger by a factor of  several
compared with the supernova remnant case at the same time, which
is crucial to the fitting of the gamma-ray spectrum of the CMZ
region at highest energy bins. As can be seen in the next
subsection (Fig.~1),  we may not get an acceptable fitting to the
spectrum of the CMZ region if $E_{p, \rm max}$ is several times
smaller. However, in the case of a hypernova explosion or some
other events which release a comparable kinetic energy ($\ga
10^{52}\,$erg) to a TDE, the result would be quite similar.}

\subsection{The gamma-ray spectra of the inner and outer regions}
{  We now look further into the  gamma-ray production during the
propagation of CRs, aiming to simultaneously reproduce the
gamma-ray spectra of both the central point-like source and the
outer CMZ region measured by HESS.}

Assuming that the cosmic-ray injection spectrum follows
$J_p(E_p,t) = C(t) E_p^{-s}\exp\left(-E_p/E_{p,\rm max}\right)$,
we can find the normalization factor $C(t)$ as
$C(t)=L_p(t)/\int_{\rm 1\,GeV}^{\infty}
E_p^{1-s}\exp\left(-E_p/E_{p,\rm max}\right)dE_p$. Since the
observed gamma-ray spectral index of HESS~J1745-290 is about
$2.1$, the value of $s$ should be around 2.2 as the cross section
of $pp-$collision increase roughly with energy as $E^{0.1}$
beyond the threshold energy. While the cosmic-ray spectral shape
follows the injected one in the inner region where particles still
propagate rectilinearly, the spectral shape in the outer region is
affected by diffusion . Given that the energy dependence of
the diffusion coefficient follows the form
\begin{equation}
D(E_p)=10^{30}\,\left(\frac{D({\rm 10\,TeV})}{10^{30}\rm cm^2/s}\right) \left(\frac{E_p}{10\,\rm TeV}\right)^\delta \,\rm cm^2/s
\end{equation}
with $\delta$ usually being in the range of $0-1$, the CR spectrum index becomes $s+\delta$ in the diffusion region. Since the gamma-ray spectral index of the CMZ region is $\sim 2.3$, we adopt $\delta=0.2$.

To find the spatial distribution of CRs, we consider a
temporally and spatially independent diffusion coefficient
$D(E_p)$, and assume that injection and diffusion have spherical symmetry. Then the
probability of finding one particle at distance $r$ away from the
source at a time $t$  after its injection is given by
\citep{Aloisio09, Dunkel07}
\begin{equation}
P(E_p,t,r)=\frac{\theta(ct-r)}{(ct)^3Z\left(\frac{c^2t}{2D(E_p)}\right)\left[1-\left(\frac{r}{ct}\right)^2\right]^2}{\rm exp}\left[-\frac{\frac{c^2t}{2D(E_p)}}{\sqrt{1-\left(\frac{r}{ct}\right)^2}} \right],
\end{equation}
where $Z(y)=4\pi K_1(y)/y$ with $K_1$ being the first-order
modified Bessel function,  and $\theta$ is the Heaviside function.
This function works for both the rectilinear propagation of
particles at $r\ll D(E_p)/c$ and the diffusive propagation at $r
\gg D(E_p)/c$, as well as for the transition regime of the two
propagation modes. We note that the {  inclusion} of both
diffusion and rectilinear regimes of propagation is a necessary
for fitting, as the spectra of the gamma-ray emissions from CMZ
and HESS~J1745-290 have different spectral indices. Indeed, the change of
spectral features from one region to another, assuming a single CR
source, indicates the changes either in the injection spectrum or
in the propagation regime of CRs. Since we assume that the index of
the injection spectrum is constant, we should consider the change in the propagation
regime. We neglect the energy losses of CRs during their
propagation via the $pp$--collision, since the energy loss time,
$t_{pp}\simeq 5\times 10^5\left(\frac{n}{100\, {\rm
cm}^{-3}}\right)^{-1}\,$yr,  is much longer than  the propagation
time to a distance of 200\,pc,  $t_{\rm diff}\simeq 10^4\
\left(\frac{r}{200\,{\rm pc}}\right)^2\left(\frac{D}{10^{30}\,{\rm
cm^2/s}}\right)^{-1}\,$~yr, unless the diffusion
coefficient is $\ll 10^{29}\,\rm cm^2/s$. { Due to the similar reason, the advection of cosmic rays
with a possible wind/outflow of a bulk velocity $v_w$ launched from the GC may be negligible given the advection timescale $t_{\rm adv}\simeq 2\times 10^5 \left(\frac{r}{200\,{\rm pc}} \right)\left(\frac{v_w}{1000\,{\rm km/s}} \right)\,$yr, which is also much longer than the diffusion timescale. }

The radial distribution of CR energy density at $t_0$ after the
{  occurrence of a } TDE can then be obtained by
\begin{equation}\label{eq:n_cr}
n_p(E_p,r,t_0)=\int\limits_{0}^{t_0} P(E_p, t_0-t, r)\,  J_p(E_p,t)\,dt,
\end{equation}
Once the spatial distribution of the protons is obtained, we can
calculate the gamma-ray  production rate in the ambient hydrogen
gas. Here we adopt the parametrized analytical formula for
gamma-ray emission of $pp-$collision given by \citep{Kelner06}
\begin{equation}\label{eq:pp}
\dot{q}_\gamma(E_\gamma,r)\equiv \frac{\dot{dN_\gamma}}{dE_\gamma}=cn(r)\int\limits_{E_\gamma}^\infty \sigma_{pp}(E_p)n_p(E_p,r)F_\gamma\left(\frac{E_\gamma}{E_p},E_p\right)\frac{dE_p}{E_p}.
\end{equation}
where $n_p(r)$ is the density of the hydrogen gas at a radial
distance $r$ to the GC, $\sigma_{pp}$ is the cross section for the
$pp-$collision and $F_\gamma$ dictates the distribution of
secondary gamma rays. For simplicity, we {  assume} two
different uniform densities for inner $15$~pc region and the outer
region respectively, while their values are chosen to fit the
measured spectral data. Then, by assuming the disk radius to be
250\,pc and the disk height to be 70\,pc, we  integrate over the
gamma-ray flux from the very center up to a projection distance of
15\,pc, and from 15\,pc up to 70\,pc to obtain the theoretical
gamma-ray flux of HESS~J1745-290 and CMZ regions respectively
(see the Appendix for details).

\begin{figure}[htbp]
\includegraphics[width=0.8\textwidth]{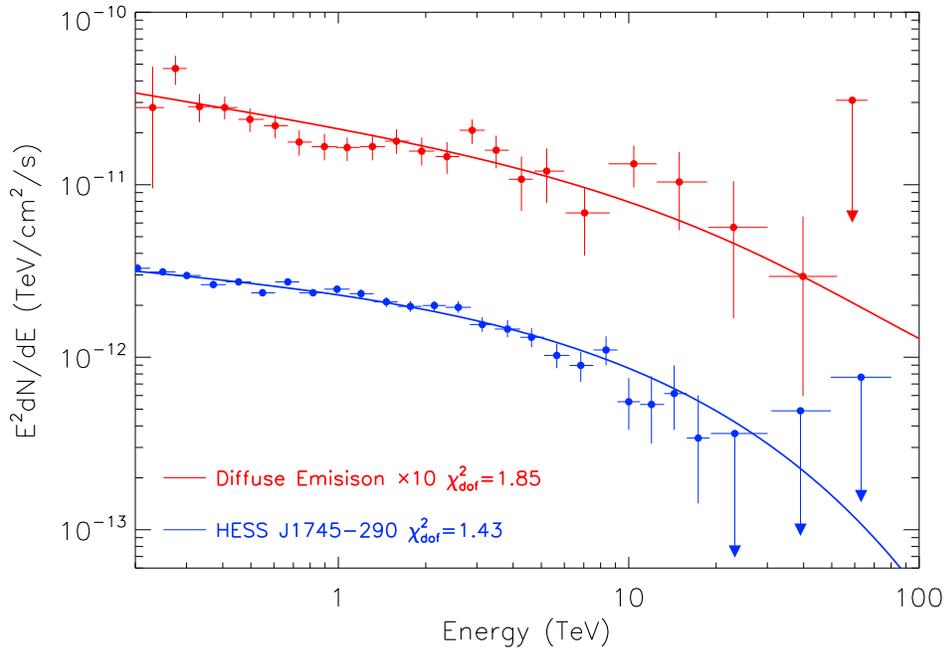}
\caption{Fitting  the gamma-ray spectra of the CMZ region (red) and HESS~J1745-290 (blue). The adopted parameters in the calculation are given in
Table.~\ref{tab:parameter}. All the data are taken from\citet{HESS_GC16}.}
\label{fig:result}
\end{figure}

In Fig.~\ref{fig:result}, fixing the injection  time
$t_0=10^4\,$yr (i.e., the TDE occurred at $10^4\,$ years ago),
{  we fit the gamma-ray spectra of both the CMZ region and
HESS~J1745-290 with the fading TDE blast wave model}. The spectrum
of the CMZ region starts to drop at $\gtrsim 10\,$TeV because the
maximum proton energy in the {  outer} region already {
starts to} fall below PeV, but the fitting is still consistent
with the HESS data at $1\sigma$ confidence level, with a reduced
chi--square of 31.4/17\footnote{ The best fitting would result in
a minimum reduced chi--square of 1 at least. Given a degree of
freedom of 17 (22 data points $-$ 5 free parameters), the
corresponding reduced chi-square for $1\sigma$ boundary should be
$>2.13$. So our fitting must be within the $1\sigma$ confidence
level although we do not obtain the best fitting here. For the
same reason we state that other fittings are also within $1\sigma$
confidence level in the text below (The 5 free parameters used in
the profile fitting are the diffusion coefficient for 10\,TeV
proton, the normalization, power--law index and the cutoff energy
of the proton injection spectrum, and the gas density. Other
parameters are either determined by the model in advance or
derived according to these parameters)}. We do not consider the
flux attenuation via $\gamma\gamma$ absorption by interstellar
background photons as it is not important in the considered energy
range\cite[e.g.,][]{Porter05, Moskalenko06, Celli16}. The fitting
to the spectrum of HESS~J1745-290 yields a reduced chi-square of
27.2/19, also at $1\sigma$ confidence level.    Technically,
we could adopt a larger acceleration efficiency $\eta$
and/or a larger explosion energy $E$ (see Eq.~\ref{eq:Emax}) to
increase the maximum energy and  thus improve the fitting to
the spectrum of the CMZ region, however, at the expense of the
 gamma-ray flux from the inner $15$~pc significantly exceeding the
$2\sigma$ upper limit  in the spectrum of HESS~J1745-290 at
20--30\,TeV. In this case, considering a clumpy distribution
of infrared photon field in the inner region would enhance the
absorption and  lead to a decrease or even a cutoff in the
spectrum of the inner region \citep[cf][]{Guo16}. Besides, the
fitting can be ameliorated by a further fine-tuning of the adopted
parameters. {  For simplicity,}  we  adopt some typical values
for the parameters, which are shown in
Table~\ref{tab:parameter}. The gas density used for the inner
15\,pc is several tens times higher than that of the outer
region. This might reflect the density difference between the core
region of {the central molecular zone} and its average value. The
radius of the blast wave {  at $t_0=10^4{\rm yr}$} is still
within 10\,pc, which is consistent with the HESS measurement. The adopted
parameters result in a total proton luminosity ($>1\,$GeV) of
about $6\times 10^{38}\rm erg/s$ at the present time. Although
this value is about three orders of magnitude higher than the
current bolometric luminosity of Sgr$~A^*$ \citep{Genzel10}, {
it reflects the past activity} of the supermassive black hole.

\begin{table}[htbp]
\centering
\begin{tabular}{llll}
\hline\hline
\multicolumn{4}{c}{Adopted Parameters \& Resulting Parameters}\\ 
\hline
$E$ & $ 1.1\times 10^{52}$erg & $L_p^{t}(>1\rm\,GeV)$  & $6.3\times 10^{38}$erg/s\\
$s$ & 2.15 & $E_{p,\rm max}^{t}$ & 150\,TeV\\
$D(10\,{\rm TeV})$ & $3.3\times 10^{30}{\rm cm^2/s}$ & $\delta$ & 0.2\\
$\eta$ & 0.25 & $\beta^t$ & $4.7\times 10^{-4}$\\
$f_{\rm CR}$ & 1.8\% & $R^t$ & 4.1\,pc\\
$n$(CMZ region) & $150\, \rm cm^{-3}$ & $n$(central region) & $5000\,\rm cm^{-3}$\\
\hline
\end{tabular}
\caption{Parameters used in Fig.~\ref{fig:result}.  Parameter with
the superscript '$t$' means the quantity is time--dependent and
the shown value is the present--time value.}\label{tab:parameter}
\end{table}

\section{Cosmic ray energy density profile in the fading accelerator scenario}
{Using the spatial distribution of the gamma-ray intensity and the
amount of target gas in the CMZ, \citet{HESS_GC16} has obtained
the projected profile of the $E\ge 10$ TeV CR energy
density up to $r\simeq 200 {\rm pc}$. Assuming a
spatially independent diffusion coefficient, \citet{HESS_GC16}
found that the data is consistent with a constant injection rate
of CRs, which is in contrast to a decreasing injection rate  in
our model. Below we study the expected cosmic ray energy density
profile from a fading accelerator and compare it with the HESS
measurements.}

\begin{figure}[htbp]
\includegraphics[width=0.8\textwidth]{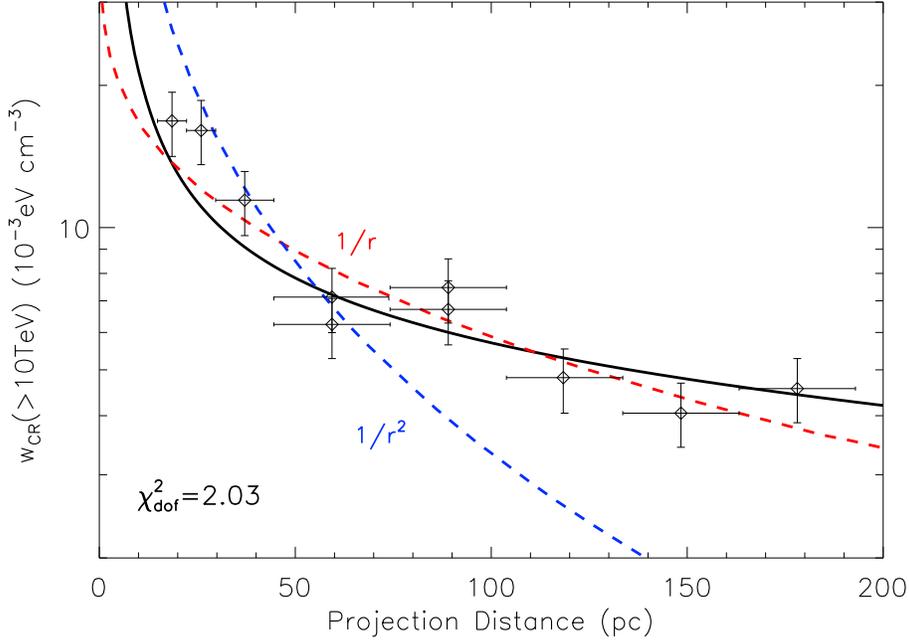}
\caption{Fitting to the measured CR energy density under the TDE blast wave model with the same parameters in Table.\ref{tab:parameter}. The black curve is calculated by averaging over the CR energy density in the line of sight for each given projection distance to the GC, yielding a reduced chi--square of 2.03. Open diamonds are measured data, the red and blue dashed lines represent the performance of the projection of a $1/r$ and a $1/r^2$ profile respectively, as shown in the Figure 2 in \citet{HESS_GC16}.}\label{fig:profile}
\end{figure}

First, we use Eq.~(\ref{eq:n_cr}) and the parameters obtained from
the spectral fitting (see Table.\ref{tab:parameter}) to calculate
the CR energy density profile in the TDE blast wave scenario.
Eq.~(\ref{eq:n_cr}) already gives the radial distribution of CR
number density. We need to integrate it over energy to get the
radial distribution of CR energy density and then average over the
line of sight to get the projected profile of CR energy density
(see Appendix for details). {  The fit to the data} is shown in
Fig.~\ref{fig:profile}, which gives a reduced chi--square of
$12.2/6$. The theoretical curve is within the $1\sigma$ deviation of the HESS data.

We note that the CR energy density profile is sensitive to CR
injection history which depends on the specific accelerator
models, while the TDE blast wave is just   one example of the
accelerators.   Meanwhile, the CR diffusion coefficient in
the CMZ region, which is not well-understood, also affects  the CR
energy density profile, as it determines  whether the
propagation mode of CRs at certain distance is rectilinear or
diffusive. So we perform a general study on  how the CR
injection history and diffusion coefficient affect the CR energy
density profile.

%
%

%

We parametrize the particle luminosity at $E_p$ as
$L(t,E_p)=L_*(t/t_*)^{-\alpha}$ with $t_*$ being the time when the
luminosity starts to follow the assumed form and $\alpha$. Then, we can write the
energy density distribution of cosmic rays at $t_0$ as
\begin{equation}\label{cr_density}
w(E_p,r)=\int\limits_{t_*}^{t_0} P(E_p,t_0-t,r)\,L(t, E_p)\,dt.
\end{equation}
We fix $t_*=0.1\,$yr in our calculation, though we note that the
obtained  density at $r$ is not sensitive to $t_{*}$ as long as
$t_0-t_{*} \gg r^2/D(E_p)$.
Following our previous assumption of a gas disk of radius 250\,pc,
we average over the CR density along the line of sight to get the
projected CR energy density profile and then compare it with
the measurement of HESS. The obtained CR energy density profile is
mainly determined by the injection time $t_0$, the power-law slope
of the injection rate history $\alpha$ and the diffusion
coefficient $D(E_p)$. In the following calculation, we will fix
$t_0$ at certain values and change the other two quantities. 
To save the calculation time, we
consider a mono-energetic injection of CRs at $E_p=10$\,TeV and calculate
the 10\,TeV CR energy density profile.
Although what HESS measures is the integrated CR energy
density above 10\,TeV, this simplification does not introduce much
error since most of the CR energies above 10\,TeV concentrate on $10\,$TeV for a
soft CR spectral index of $2.4$ at the CMZ region.

\begin{figure}[htbp]
\includegraphics[width=0.45\textwidth]{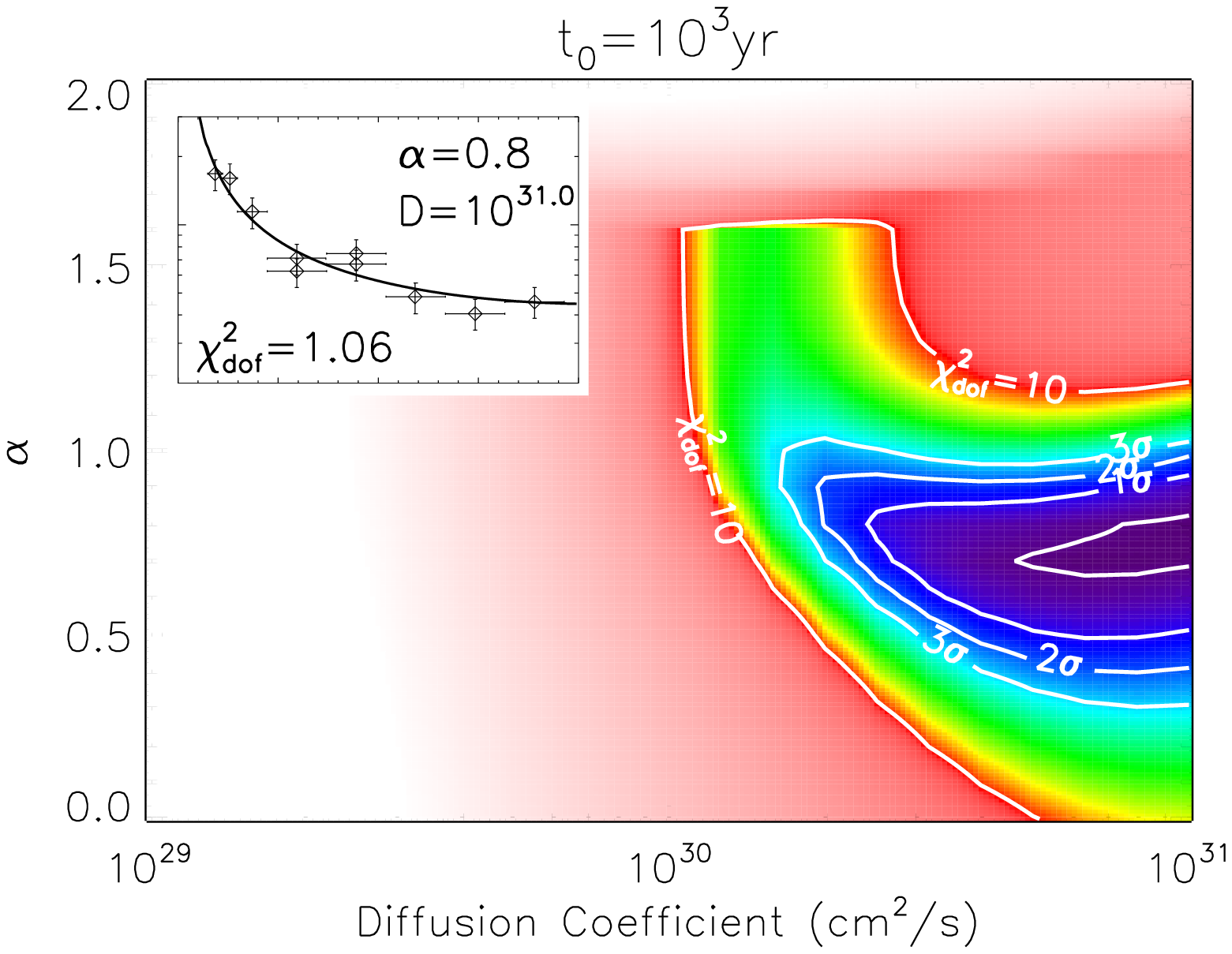}
\includegraphics[width=0.45\textwidth]{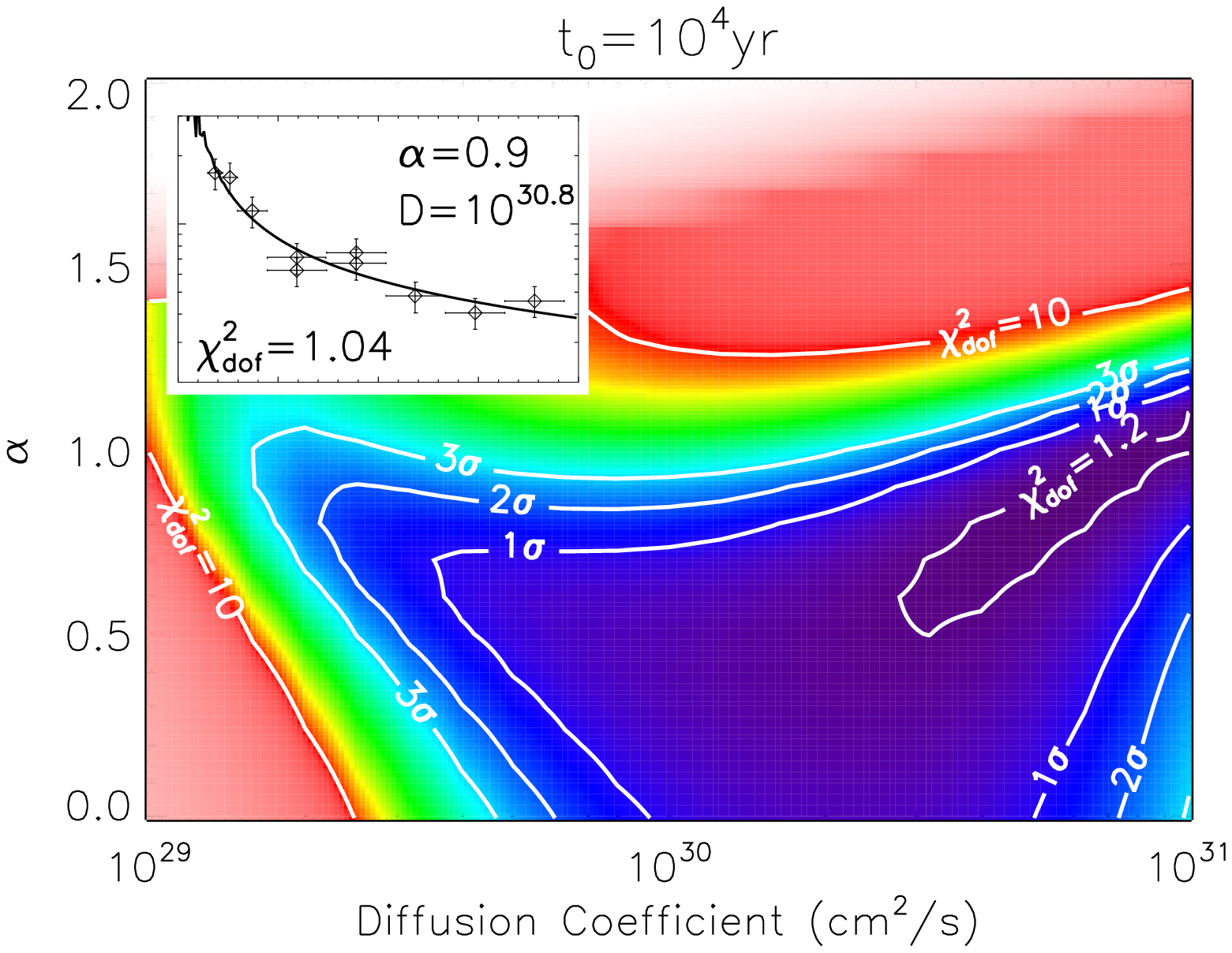}\\
\includegraphics[width=0.45\textwidth]{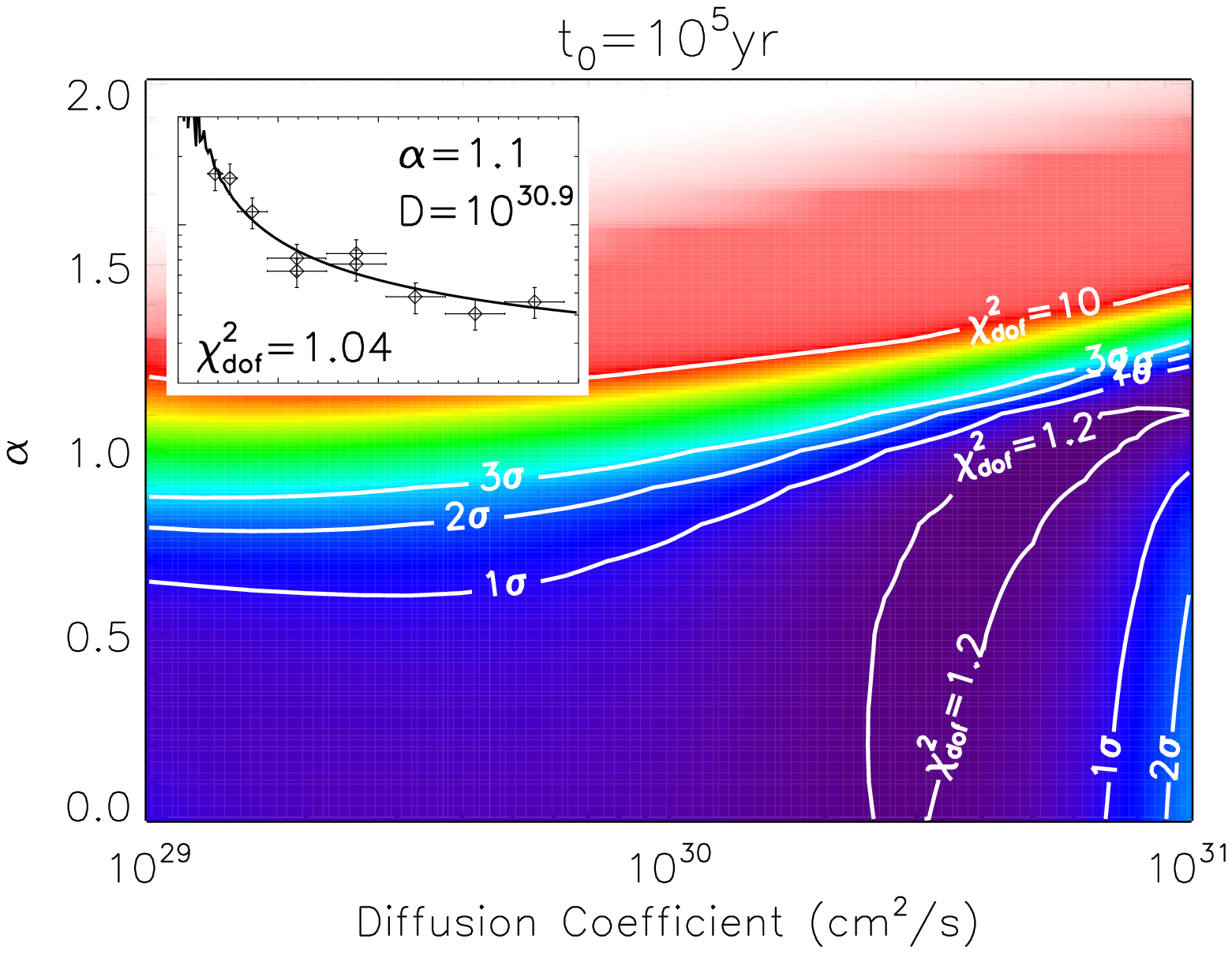}
\includegraphics[width=0.45\textwidth]{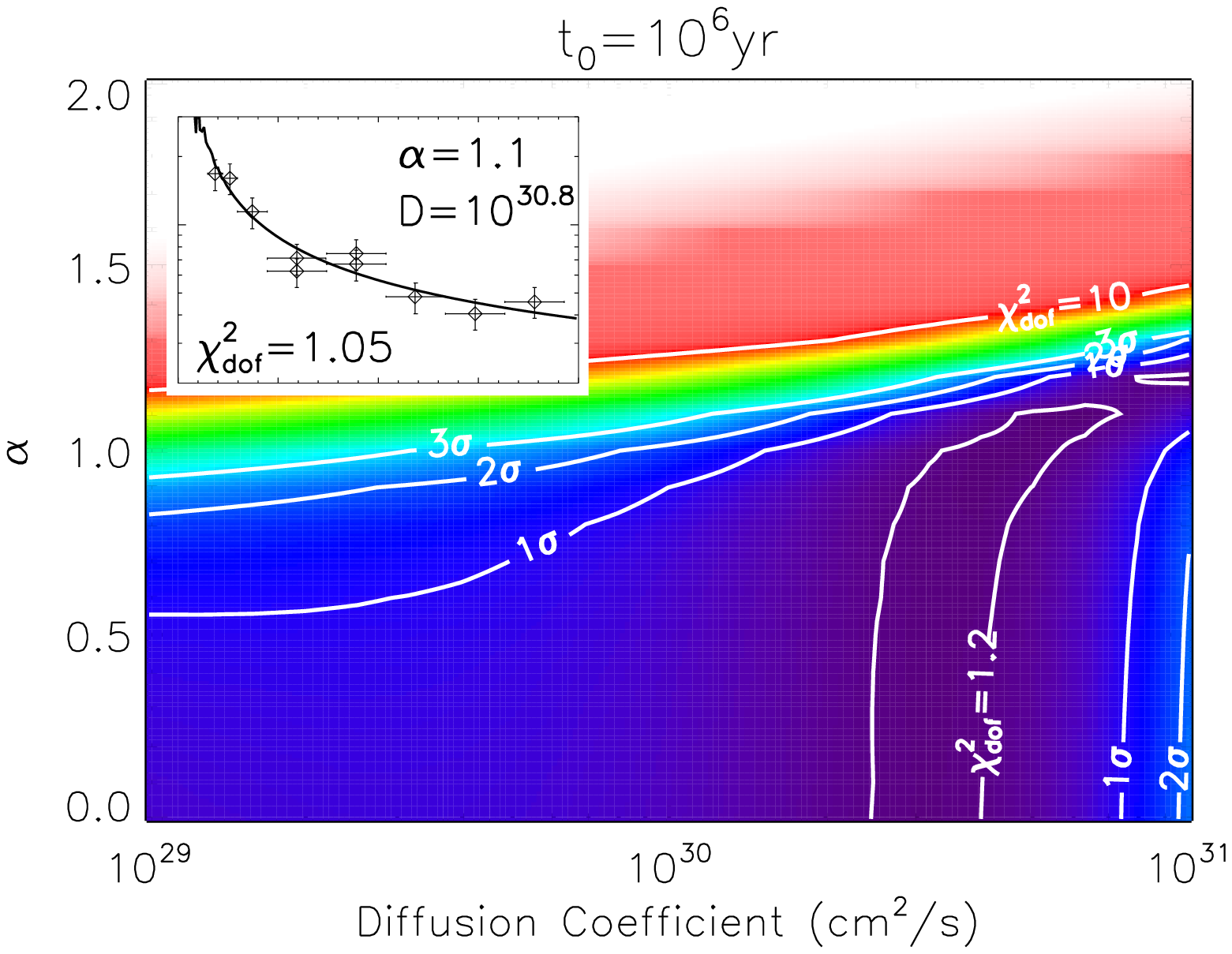}\\
\includegraphics[width=0.9\textwidth]{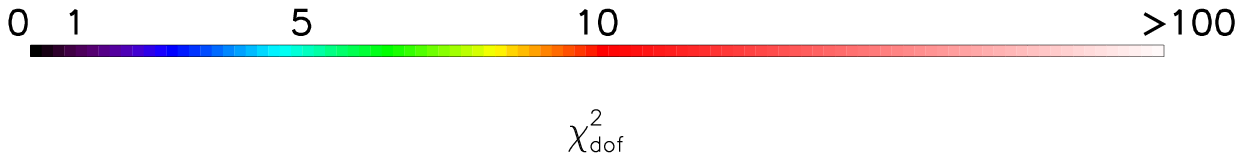}
\caption{The reduced chi--square for different combinations of injection history $\alpha$ and diffusion coefficient $D$ for 10\,TeV protons given different injection time. Different colors corresponds to different values of the reduced chi--square, as shown in the color bar in the bottom. Note that the color changes linearly in the range of $\chi^2_{\rm dof}=0-10$ and logarithmically in the range of $10-100$. The 1$\sigma$, 2$\sigma$, 3$\sigma$ confidence region as well as the contours for $\chi_{\rm dof}=1.2, ~10$ are marked in the figure. The inset figures show the best fits to the measured CR energy density in the considered range of $\alpha$ and $D$.}
\label{fig:chi_sq}
\end{figure}

The results with different injection time of $t_0=10^3\,$yr,
$10^4\,$yr, $10^5\,$yr and $10^6\,$yr are shown in different
panels in Fig.~\ref{fig:chi_sq}. The reduced chi-square test
is used to evaluate the goodness of fit. Different colors represent the
corresponding values (see the color bar) of the reduced chi-square
($\chi^2_{\rm dof}$) for each combination of $\alpha$ and $D$.
$1\sigma$, $2\sigma$, $3\sigma$ confidence regions are marked in
the figure, as well as the contours for $\chi^2_{\rm}=1.2, 10$ for
comparison. Since $L_*$ only relates to the amplitude of the
energy density profile, we do not concern ourselves with it and
its value is chosen to get the minimum chi-square for the given
values of $\alpha$ and $D$. The inset figure in each panel shows
the fit to the measured CR energy density profile with the
best-fit parameters for a given $t_0$. We find that a
decreasing injection rate with $\alpha\sim 0.5-1.0$ and a large
diffusion coefficient $\gtrsim 10^{30.5}\,\rm cm^2/s$ for 10\,TeV
protons can give good fitting to the measured data. We can also see the deviation from the data in the TDE
blast wave model, in which $t_0=10^4\,$yr, $\alpha=0.8$ and $D(10\,{\rm TeV})=3.3\times 10^{30}\,\rm cm^2/s$,
is $<1\sigma$, being consistent with the result in Fig.\ref{fig:profile}.

The decreasing
injection rate of a fading source leads to the flattening of the
spatial distribution of CR density, compared to the measured
$\sim 1/r$ dependence which arises from a constant injection with a spatially
independent diffusion coefficient. However, a large diffusion
coefficient results in a transition to the $\sim 1/r^2$ dependence
for the rectilinear regime at small distances, which alleviates
this problem, leading to a reasonable fit to the data. Actually,
for a small $t_0$ and a small $D$, constant injection does not
give a very good fitting to the data. This is because some particles
have not diffused to large $r$ for a finite injection
time $t_0$ and, hence, the profile can not keep the $1/r$ form
\footnote{Mathematically, in the purely diffusion regime, a
constant injection rate $L$ leads to a distribution of energy
density as $\frac{L}{4\pi Dr}{\rm
erfc}\left(\frac{r}{\sqrt{4Dt}}\right)$ \citep{Atoyan95}. The
profile starts to drop when $r$ approaches $\sqrt{4Dt}$.}. We note
that the favored diffusion coefficient in the fit is about
one order of magnitude larger than that in the Galactic disk at
the same energy, which is $\sim 10^{28}{\rm \, cm^2/s}\,
(E/1\,\rm GeV)^{1/3}$ \citep[e.g.][]{Strong10}. However, the diffusion coefficient in the
CMZ region is unclear currently. Indeed, due to the different
environment at the Galactic Center and the Galactic disk,
the diffusion coefficients in these two regions may be different.
We will further discuss the diffusion coefficient in the next section.

\section{Discussions and Conclusions}
In this work, we proposed a fading accelerator located in the
inner $10$~pc of the GC as the source for the gamma-ray emission
of both the point-like source HESS~J1745-290 and the CMZ region.
Given the decreasing maximum acceleration energy in the injected
proton spectrum, gamma rays of the inner region presents a cutoff
around $10$~TeV as they are produced via $pp-$collision by more
recently injected protons that have a lower maximum energy, while
the gamma rays of the outer CMZ region arise from earlier injected
protons that have a larger maximum energy and hence produce a
spectrum extending to energies beyond tens of TeV. The fading
accelerator could be due to some past activity of the
supermassive black hole, such as a blast wave driven by  a TDE
{due to the supermassive black hole in the GC}. Following the
Sedov solution
 of the blast wave, we obtained the
evolution of the particle injection rates and maximum acceleration
energy in such a scenario. A simultaneous fit to the spectra of
both HESS~J1745-290 and the  CMZ region is  obtained within
$1\sigma$ deviation from the HESS data. We find that the measured
CR energy density profile can also be reproduced with the same
parameters. Furthermore, we performed a general study of the
effects of the CR injection history  and the diffusion coefficient
on the resultant energy density profile and found that the
fading accelerator can reproduce the measured data for different
injection times.

It may be worth noting the possibility that gamma rays in
the inner region may have a  different origin from that in
the outer CMZ region. {In this case, the fading accelerator
model can still account for the gamma-ray emission in the CMZ
region and the fit to the data could be improved as there is no
constraints from the inner source anymore}. We may adopt a larger
kinetic energy and/or a higher acceleration efficiency to get a
higher $E_{\rm max}$, so that the calculated spectrum in the CMZ
region can extend up to tens of TeV without softening. In the
meanwhile, we need to adopt a lower hydrogen gas density for the
inner region in order not to overshoot the observed gamma-ray flux
of HESS~J1745-290.

As we already mentioned in the last section, to obtain a good fit
to the CR energy density profile, the favored diffusion
coefficient in the CMZ region is about one order of magnitude
larger than that in the Galactic disk. Such a large diffusion
coefficient would lead to a purely rectilinear propagation regime
of CRs inside the inner $\ll D/c \sim 30$\,pc, which also implies that CRs are
still not fully isotropized even at tens of pc
\citep{Prosekin15}. The gamma-ray fluxes from such regions are
enhanced (reduced) if the  most CRs are moving towards (away
from) our line of sight. This is because the emitted gamma--rays
will concentrate on the moving direction of their
ultra-relativistic parent protons (i.e., the beaming effect). As a
result, we may underestimate the gamma-ray flux from the
HESS~J1745-290 while overestimate the gamma-ray flux from the CMZ
region, as we have assumed CRs are isotropized in our
calculation. However, we note that the theoretical flux of
HESS~J1745-290 is barely affected by this effect. This is because,
given a much higher gas density in the inner 15\,pc region than
the outer region,  the flux of HESS~J1745-290 should be dominated
by the emission from the inner 15\,pc region, which  has
spherical symmetry in our model. So the enhanced flux is exactly
balanced by the shrunk observable region. The extent of the
overestimation on the flux of the CMZ region depends on how far
the real distribution of CRs deviate from the isotropic one. Since
the pitch angle distribution of CRs becomes broader as they
propagate, we may expect a typical angular size of a bundle of CRs
after propagation a distance of $r$ as \citep{Prosekin15}
\begin{equation}
\theta=\sqrt{\frac{2r}{3D/c}}.
\end{equation}
Substituting the  obtained parameters into the above
equation, we find that the angular  size of 100\,TeV protons is
about { $30^\circ$ at 20\,pc and $50^\circ$ at 70\,pc}, implying a
considerable fraction of CRs that initially propagate away
from us are  already deflected to our direction. Besides,
CRs are already isotropized at larger radius. So we may
expect that the overestimation of the flux from the CMZ region is
not severe and can be modulated by tuning the gas density in the
CMZ region.

To simplify the calculation, several assumptions have been made in
our calculations,  such as a uniform gas density, a spatially
independent and isotropic diffusion coefficient, a temporally
independent particle acceleration and injection efficiency, and
etc. We note that the large diffusion coefficient may be avoided
if more complicated factors are taken into account. For
example, if the diffusion coefficient is spatially dependent, such
as increasing outwardly, a smaller diffusion coefficient might
also work. This is due to that a faster diffusion (i.e., larger
$D$) leads to a smaller density, so an outwardly increasing
diffusion coefficient could yield a steeper profile of CR energy
density compared with {that in the case of} a spatially independent
diffusion coefficient, and hence fit the data with a smaller
diffusion coefficient at the GC. Besides, if
we consider the capability of { the confinement of CRs by the
blast wave} decreases with time (i.e., $f_{\rm CR}$ increases with
time), the injection rate would decrease less rapidly with time,
and then a smaller diffusion coefficient could also work {
well} in fitting the CR density profile. More realistic modeling
using complicated numerical calculations/simulations on the
relevant processes would be useful for a more careful study. We
leave this study to a future publication.

In the present model, cosmic rays are injected by a fading
accelerator,  probably arising from { some types of explosion
events}, such as a { TDE}. Provided that the event rate of the
explosion is  $10^{-5}-10^{-4}\,{\rm yr^{-1}}$, such a fading
accelerator would  occur repeatedly on the timescale of
$>10^{5}$yr. In this sense, the particle injection rate appears to
be quasi--constant on such long timescales. {Indeed, there may be} no
strict constant injection in nature. Every source {may be}
variable at different time scales. We note that the measured
gamma-ray emission in the CMZ region probes the CR injection on a
time scale of $10^3-10^4$\,yr or longer, while the gamma-ray from
the inner 10\,pc mainly reflects the CR injection  in recent
$10-100$\,yr. Thus, one alternative scenario is that the source
(quasi-)constantly injected CRs (i.e., $\alpha=0$) with a spectrum
extending up to $>$PeV during the past ten thousand years or
longer while the source is variable on  a short timescale of
$10-100\,$yr. Coincidentally, it is just in a less
active period during the recent ten years, resulting in a low
particle acceleration efficiency and, subsequently, the cutoff
in the measured spectrum of the inner region. A
further study on the CR acceleration and injection is required to
reveal more details in this scenario, which, however, is beyond
the scope of this work.

\acknowledgments {We thank Felix Aharonian for the valuable
comments and suggestions, and thank Tony Bell for the helpful
discussions. This work is supported by the
National Basic Research Program (973 Program) of China
under Grant No.~2014CB845800,  the National Natural Science
Foundation of China  under  Grants  No.~11273016 and No.~11625312.}

\appendix
\section{Average/Integration over line of sight}\label{appendix}
\begin{figure}[htbp]
\includegraphics[width=0.8\textwidth]{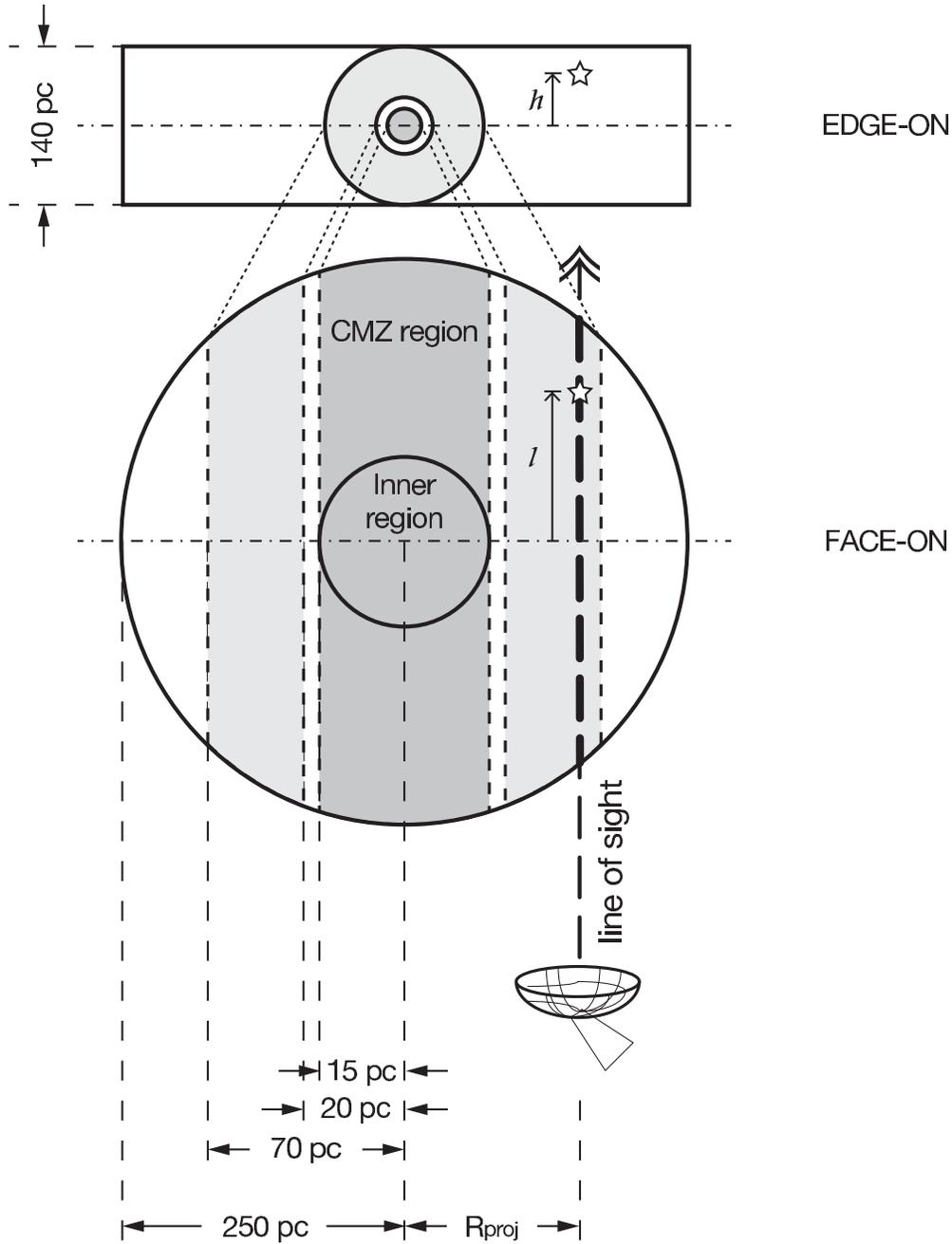}
\caption{A sketch for the geometry of the gas disk at the GC used in our calculation. The darker and lighter hatched regions are the spectrum extracting region of HESS~J1745-290 and the CMZ region respectively.}\label{fig:sketch}
\end{figure}

The geometry of the central molecular zone including the inner region in our model is shown as Fig.~\ref{fig:sketch}. We assume the outermost radius of the disk is $R_d=250\,$pc and the thickness of the disk is 140\,pc (or with a height $H_{\rm max}=70\,$pc).
The line--of--sight average CR density in the disk plane is given by
\begin{equation}
\bar{w}_{CR}(>10{\rm \,TeV}, R_{\rm proj})=\frac{1}{l_{\rm max}}\int_{0}^{l_{\rm max}}w(>10\,{\rm TeV},r)dl
\end{equation}
where $l_{\rm max}=\sqrt{R_d^2-R_{\rm proj}^2}$ and $r=\sqrt{l^2+R_{\rm proj}^2}$.

The line--of--sight integrated gamma--ray flux from the center ($r=0$) to a circle with a projected radius of $R_{\rm proj}$ is
\begin{equation}
F_\gamma(E;0, R_{\rm proj})=\frac{8}{4\pi d^2}\int_0^{R_{\rm proj}}dh \int_0^{r_{\rm proj, max}(h)} dr_{\rm proj}  \int_0^{{l_{\rm max}}}\dot{q}_\gamma (r, E)dl
\end{equation}
where $q_\gamma(r,E)$ is the gamma-ray emissivity in unit volume at energy $E$ at a radial distance $r=\sqrt{l^2+r_{\rm proj}^2+h^2}$ to the center. $d=8.5\,$kpc is the distance between the GC and the Earth. The upper limit in the integration $r_{\rm proj, max}(h)=\sqrt{R_{\rm proj}^2-h^2}$ and $l_{\rm max}=\sqrt{r_{\rm proj, max}^2-r_{\rm proj}^2}$.  The factor 8 arises from the fact that the above integration actually counts only 1/8 of the total emission of the considered region given the symmetry in the distribution gamma-ray emissivity density.
To obtain the line--of--sight integrated gamma--ray flux within a ring from a projected radius of $R_{\rm proj, min}$ to $R_{\rm proj, max}$, we need to perform the calculation $F_\gamma(E;R_{\rm proj, min}, R_{\rm proj, max})=F_\gamma(E;0, R_{\rm proj, max})-F_\gamma(E;0, R_{\rm proj, min})$.

\bibliographystyle{aasjournal}
\bibliography{ms_revised}

\end{document}